# Bidirectional Photoinduced Carrier Transfer in Fluorinated Quasi-2D Perovskites Governing Enhanced Photocurrent Generation

*Soumya Halder [1,2†], Koushik Gayen [1,2†], Nagendra S. Kamath [1,2] and Suman Kalyan Pal [1,2*]*

[1]School of Physical Sciences, Indian Institute of Technology Mandi, Kamand, Mandi-175005, Himachal Pradesh, India

[2]Advanced Materials Research Centre, Indian Institute of Technology Mandi, Kamand, Mandi-175005, Himachal Pradesh, India

[†]These authors contributed equally to this work.

AUTHOR INFORMATION

**Corresponding Author**

*E-mail: suman@iitmandi.ac.in; Phone: +91 1905267040

**Abstract**

Quasi-two-dimensional (quasi-2D) metal halide perovskites exhibit rich phase heterogeneity that profoundly influences light-matter interactions and charge transport. However, the fundamental mechanisms governing carrier transfer across distinct phases remain poorly understood. Here, we demonstrate effective electron-hole separation in fluorinated multilayered quasi-2D perovskite films nominally prepared for three layers, using femtosecond transient absorption spectroscopy. The films are revealed to comprise a heterogeneous phase distribution (with 1, 2, 3 layers and bulk) naturally stacked along the growth direction. Our ultraviolet photoelectron spectroscopy (UPS) measurements, show the type-two band alignment between the small-n (layer number) phases and the bulk. This alignment drives charge separation via both direct and sequential carrier transfer mechanisms, whereby electrons preferentially migrate into the bulk domains while holes accumulate in the small-n layers, extending even to single layer phase-a process only rarely observed in previous studies. The nearly symmetric transfer times of electrons and holes yield an efficient and balanced spatial separation of carriers. Global target analysis employing a carrier transfer model quantitatively reproduces the spectral evolution, providing a rigorous validation of the mechanism. Nonetheless, we found photocurrent enhancement in the diode devices of this quasi-2D perovskite as a consequence of the efficient transfer of photocarriers in the opposite directions. This work delivers a comprehensive picture of interphase charge transfer in fluorinated quasi-2D perovskites and highlights strategies to engineer directional separation pathways for high-performance photovoltaic, optoelectronic, and quantum devices.

1. Introduction

Halide perovskites exhibit exceptional optoelectronic properties-including high absorption coefficients[1-5], high carrier mobilities[6-8], long carrier diffusion lengths[9-11], and remarkable defect tolerance[12-16]- together with pronounced nonlinear optical responses[17-19], strong exciton–carrier interactions[20, 21], and efficient carrier-phonon coupling[22-24]. These synergistic features have fueled their rapid rise across photovoltaics, light-emitting diodes (LEDs), photodetectors, lasers, and restive switching memory devices [25-29]. Single-junction perovskite solar cells (PSCs) have already surpassed 26% power conversion efficiency, placing them among the most efficient thin-film technologies[30, 31]. Yet, the commercialization of three-dimensional (3D) perovskites remains hampered by their intrinsic environmental instability. The soft ionic lattice and weak noncovalent interactions between organic cations (e.g., $MA^+$, $FA^+$) and the inorganic framework render these materials highly vulnerable to oxygen and moisture[32, 33], necessitating strategies that enhance stability without compromising electronic performance. Interface engineering and surface passivation with two-dimensional (2D) layers has demonstrated improved durability[34, 35], underscoring the promise of hybrid 3D/2D architecture.

Building on this concept, quasi-two-dimensional (quasi-2D) Ruddlesden–Popper perovskites have emerged as a particularly attractive class. In these materials, bulky organic spacers form layered structures that shield the lead–halide framework from environmental stressors[36-38] while simultaneously enabling precise control over bandgap and exciton binding energy. Their general formula, $L_2A_{n-1}B_nX_{3n+1}$, naturally yields quantum-well heterostructures whose optical and electronic properties depend on the quantum-well thickness (n). By tuning n, one can engineer excitonic and band-edge properties with molecular precision[39-41]. Importantly, because solution-processed films inherently contain both low-n and bulk-like domains, quasi-2D perovskites act as self-assembled 2D/3D heterostructures[1]. This built-in heterojunction can provide intrinsic carrier and energy transfer pathways[42, 43], while the layered architecture imparts enhanced stability[44]. As a result, quasi-2D perovskites stand out as versatile and robust candidates for next-generation photovoltaics, LEDs[45, 46].

Nevertheless, the unavoidable multiphase character of quasi-2D films produces a complex energetic landscape in which exciton funneling[47], carrier separation via charge transfer, and exciton recombination[48] can take place. A central unresolved question concerns the underlying band alignment: type-I alignment would favor exciton funneling[49], whereas type-II alignment enables interphase charge transfer and long-range carrier separation, critical for efficient photovoltaics[50]. While ultrafast electron transfer from high-bandgap (low-n) to bulk-like domains has been widely reported, the reported timescales vary dramatically (from <100 fs to >1 ns)[51]. In contrast, direct evidence for hole transfer from bulk to low-n (n<2) phases has remained elusive[42, 52, 53], hindered by weak experimental signatures and the sensitivity limits of conventional ultrafast probes.

Here, we investigate photoinduced exciton/carrier dynamics on layered (4-FBA)$_2$(MA)$_{n-1}$Pb$_n$I$_{3n+1}$ films with nominal n=3 by combining photoluminescence excitation (PLE) and femtosecond transient absorption (TA) spectroscopy. PLE unambiguously rules out exciton funneling[54], establishing interphase charge transfer as the dominant relaxation pathway. Structural analysis reveals vertical phase stratification, with low-n domains near the substrate and bulk-like domains enriched at the film surface, creating a natural directional energy landscape. By employing complementary excitation geometries-back-side excitation at 480 nm and front-side excitation at 750 nm-under fluences well below the threshold <10 μJ cm$^{-2}$ per pulse for nonlinear many-body effects[53, 55], we selectively triggered electron transfer into bulk-like domains and hole transfer into n=1 domains. This dual-geometry approach uncovers striking bidirectionality: electrons photogenerated in small-n domains transfer directly and sequentially into bulk-like phases, while holes generated in the bulk migrate efficiently into low-n domains. Both processes unfold on ultrafast timescales (hundreds of femtoseconds to tens of picoseconds), producing long-range spatial separation of carriers across the layered stack.

Crucially, fluorination of the organic spacer amplifies these transfer processes by enhancing interfacial coupling and dielectric screening. This results in stronger electron injection into bulk domains and robust hole stabilization in n=1 phase. Our findings provide the direct spectroscopic evidence of efficient hole transfer into the lowest-n phase in quasi-2D perovskites, while also revealing a pronounced excitation-energy dependence of electron transfer: high-energy excitation above the n = 1 band edge drives direct transfer into the bulk, whereas band-edge excitation of individual phases gives rise to sequential transfer pathways. Beyond clarifying this mechanistic distinction, we establish that fluorinated quasi-2D perovskites can function simultaneously as carrier generator and extractor for device applications, eliminating the need for additional charge transport layers. This unique bidirectional charge-separation mechanism establishes a new paradigm for the design of robust, high-performance perovskite optoelectronic devices.

## 2. Results and Discussion

**2.1 Film Characterization**

The 2D perovskite films were prepared by spin-coating and subsequent thermal annealing (see Note S1: Experimental Methods in Supporting Information (SI) for details). SEM images reveal uniform layered structure, with film thickness of ~300 nm (Figure S1a-b). The film surface morphology examined by AFM shows a smooth and uniform texture with an average roughness of 11.6 nm (Figure S1c), confirming good-quality film formation. Figure 1a presents the UV-vis absorption spectrum of the 2D perovskite film, which reveals a series of well-defined excitonic resonances. Three prominent absorption peaks are observed at 520 nm (2.38 eV), 570 nm (2.18 eV), and 614 nm (2.02 eV), corresponding to the n=1, n=2, and n=3 phases, respectively. In addition to these higher-energy excitonic features, a distinct peak appears at 752 nm (1.65 eV), which can be attributed to perovskite domains with much larger n values, approaching the bulk (3D) limit. The coexistence of these well-resolved absorption bands provides direct evidence for the formation of multiple layered phases within the film. Interestingly, despite employing a precursor solution designed to promote preferential growth of the n=3 phase, the optical signatures clearly indicate the presence of a heterogeneous phase distribution. Such phase mixing is frequently encountered in solution-processed 2D perovskites, where kinetic limitations and competing crystallization pathways hinder complete phase purity[23]. The mixed phase distribution arises because the local molar ratio of 4-FBA insulating cations to MAI and $PbI_2$ precursors dynamically changes during solvent evaporation, thereby driving the nucleation of domains with different n values[42].

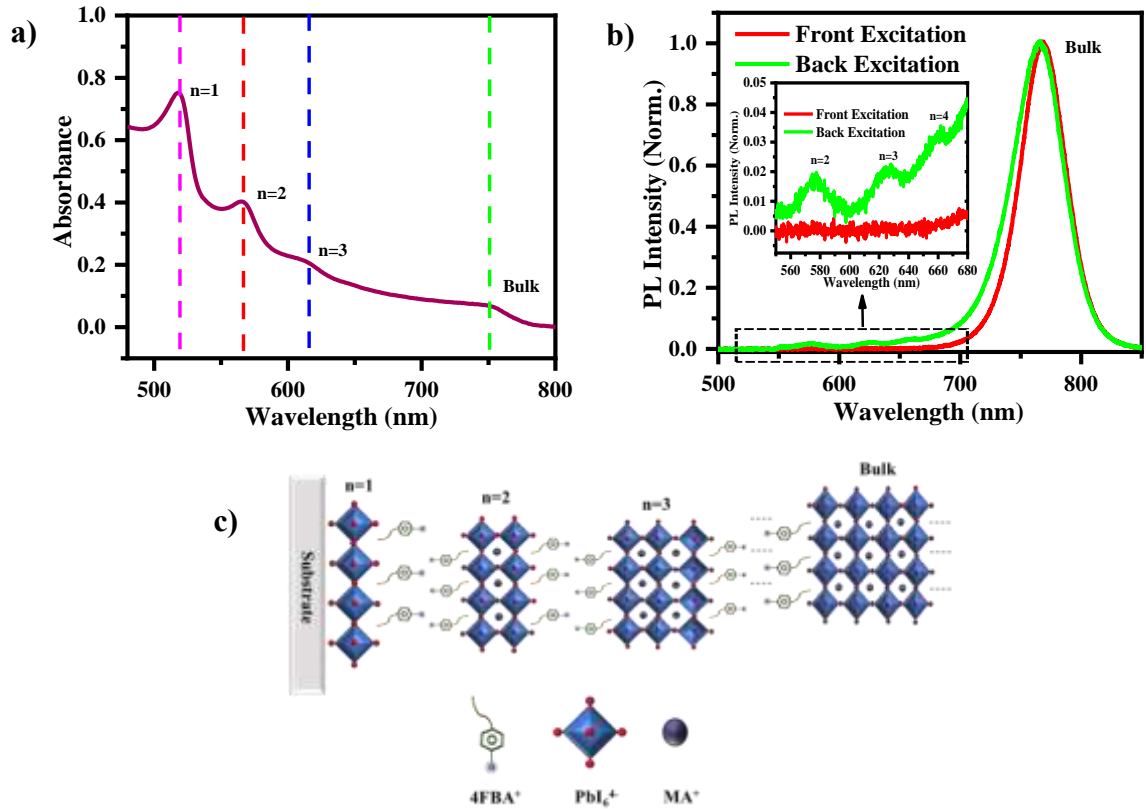

**Figure 1.** Structural and optical characterization of the 2D perovskite film. (a) UVvis absorption spectrum of the (4-FBA)$_2$(MA)$_{n-1}$Pb$_n$I$_{3n+1}$ film prepared with a nominal n = 3 composition, revealing a series of well-defined excitonic resonances. Distinct absorption peaks are observed at 520 nm (2.38 eV), 570 nm (2.18 eV), and 614 nm (2.02 eV), corresponding to the n = 1, n = 2, and n = 3 phases, respectively, along with a pronounced peak at 752 nm (1.65 eV) arising from higher-n domains approaching the bulk (3D) limit. These features confirm the coexistence of multiple layered phases despite precursor conditions optimized for n = 3. (b) Steady-state PL spectra recorded under front-side and back-side excitation. The main panel displays the dominant bulk-like emission observed for both excitation configurations, whereas the inset highlights the additional emission features from the n=2, n=3, and n=4 phases that appear under back-side excitation, evidencing the heterogeneous phase distribution within the film. (c) Schematic illustration of vertically aligned perovskite domains spanning from n = 1 to bulk-like phases, forming a phase-gradient architecture across the film thickness.

To further probe the spatial distribution of the coexisting phases, steady-state photoluminescence (PL) measurements were performed under both front-side (excitation from the film surface) and back-side (excitation through the substrate) geometries. In both cases, a dominant emission peak centered at 765 nm is observed, consistent Schematic illustration of vertically aligned perovskite domains spanning from n=1 to bulk-like phases, forming a phase-gradient architecture across the film thickness. with the emission due to radiative recombination from the bulk-like phase (Figure 1b). Notably, under back-side excitation, additional emission features emerge from lower-n perovskite phases. The disparity between the front- and back-excitation spectra indicates a vertical phase gradient across the film thickness, wherein lower-n phases preferentially reside near the substrate, while bulk-like domains accumulate toward the film-air interface (Figure 1c).

The structural characteristics of the film were further examined by X-ray diffraction (XRD) analysis (Figure S2). The XRD pattern reveals a highly crystalline material with intense reflections at 15° and 29°, indexed to the (011) and (022) planes of the bulk-like phase. In addition, distinct low-angle reflections at 7.3° and 13.8° are observed, signifying the presence of lower-n layered phases[56]. The coexistence of these high- and low-angle reflections corroborates the spectroscopic evidence (UV-vis and PL) for a mixed-phase composition encompassing a wide distribution of n values. Importantly, the pronounced intensity of the (011) and (022) reflections suggest a preferential vertical orientation of the $[PbI_6]^{4-}$ octahedral layers, aligned perpendicular to the substrate surface[57]. This vertical alignment is consistent with the PL findings (Figure 1b), which revealed a phase-gradient architecture across the film thickness, with lower-n phases enriched near the substrate and bulk-like domains dominating at the film surface. Together, these structural and optical characterizations support the schematic representation in Figure 1c, in which quasi-2D perovskite layers are vertically stacked to form a graded phase distribution throughout the film.

## 2.2 Band Alignment

To know the positions of valance band maxima (VBM) and conduction band minima (CBM) across different phases of $(4\text{-FBA})_2(MA)_{n-1}Pb_nI_{3n+1}$ perovskite films, ultraviolet photoelectron spectroscopy (UPS) measurements was performed. using monochromatic He-I source with a radiation energy, $E_{He(I)} = 21.2$ eV. UPS spectra for the different phases of the 2D perovskites are presented in Figure S3. The secondary electron cutoff ($E_{cutoff}$) values (Figure S4), were used to calculate the corresponding work functions ($E_{He(I)}-E_{cutoff}$) which were found to be 4.95, 4.89, 4.84, and 4.80 eV for n=1, n=2, n=3 and bulk respectively. The valance band onset energies ($E_{onset}$) were determined by extrapolating the linear region of the leading edge of the valence band spectra (Figure S5), yielding values of 1.06, 1.17, 1.31, and 1.46 eV for n=1, n=2, n=3 and bulk respectively. The corresponding valance band maximum ($E_{VBM}$) and conduction band minimum ($E_{CBM}$) energies were estimated using the following relations

$$E_{VBM} = -\left(E_{He(I)} - E_{cutoff} + E_{onset}\right) \quad (1)$$

$$E_{CBM} = E_{VBM} + E_{bandgap} \quad (2)$$

Using the bandgap values (2.31, 2.10, 1.89, and 1.57 eV for n=1, 2, 3, and bulk, respectively) extracted from the absorption spectrum, the absolute band positions were obtained and are presented in Figure 2a. The results reveal a systematic downward shift of both the valence and conduction band edges with increasing n, signifying a progressive reduction in quantum confinement from the lower-n phases toward the bulk.

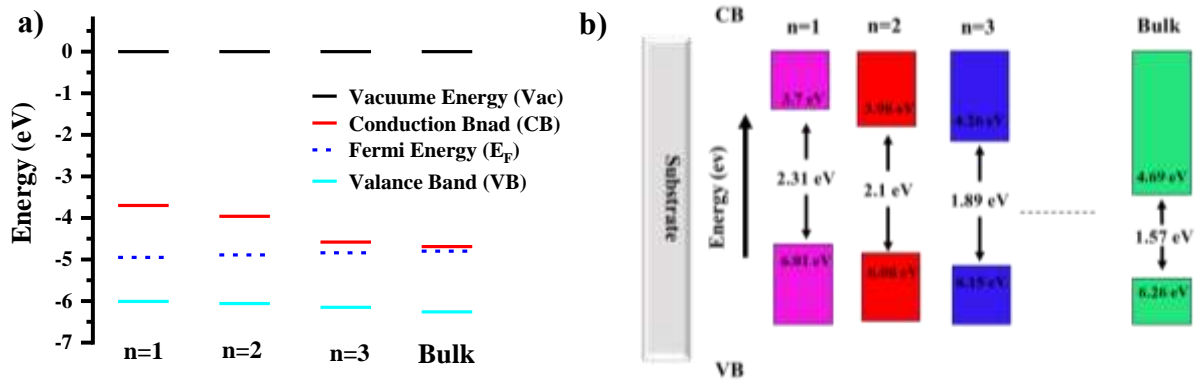

**Figure 2**. (a) Conduction and valence band edge positions of the different phases (n=1, 2, 3, and bulk) of $(4\text{-FBA})_2(MA)_{n-1}Pb_nI_{3n+1}$ films determined from UPS measurements. (b) Schematic illustration of the interphase band alignment, revealing a type-II staggered configuration among the layered perovskite phases.

Based on the experimentally derived band-edge positions, the band alignment diagram (Figure 2b) illustrates a clear type-II staggered heterostructure across the different n phases. The CBM of the bulk phase lies significantly lower than those of the small-n phases (n=1-3), while the VBMs are slightly higher relative to the bulk. This gradual convergence of the CBM and VBM levels toward the bulk phase with increasing n establishes a natural energetic gradient that facilitates photoinduced electron transfer from the small-n to bulk phases and hole transfer in the opposite direction.

## 2.3 Carrier Transfer Dynamics

### 2.3.1 Under Back Excitation

To directly unravel the interphase carrier transfer pathways in mixed-n perovskite films, femtosecond transient absorption (TA) spectroscopy was employed, providing ultrafast temporal resolution of the carrier dynamics. Excitation with a 480 nm pump (fluence ~ 0.3 μJ cm$^{-2}$) in a back-excitation geometry, (Figure 3a), where excitation predominantly addressed the small-n phases near the glass interface, the TA spectra (Figure 3b) revealed four distinct photo-bleach (PB) features at 520, 570, 624, and 778 nm, corresponding to the excitonic resonances of the n=1, 2, 3, and bulk phases, respectively. These PB signatures arise from state filling of the associated electronic transitions, accompanied by broadband photoinduced absorption (PIA) spanning both sub- and above-bandgap regions, consistent with excited-state absorption and bandgap renormalization[58]. The temporal evolution of the bleach features exposes a strikingly phase-dependent behavior. The PB intensities of the n=1-3 phases undergo a rapid decay, while the bulk PB at 778 nm grows concurrently (Figure 3b). The observed anticorrelation could arises due to transfer of carriers from lower-n phases to the bulk. However, it could be consistent with excitonic (singlet) energy transfer, which can be mediated by spectral overlap between the emission of the small-n phases and the absorption of the bulk. Kinetic traces extracted at the excitonic positions of each phase (Figure 3c) further corroborate this scenario: the n=1-3 bleaches decay within ~200 fs, nearly synchronously, while the bulk bleach exhibits a delayed rise, reaching maximum population at ~17 ps. Biexponential fits of these dynamics (Table S1, SI) quantify this interphase charge redistribution with characteristic lifetimes consistent with ultrafast carrier or energy transfer and accumulation in the bulk phase.

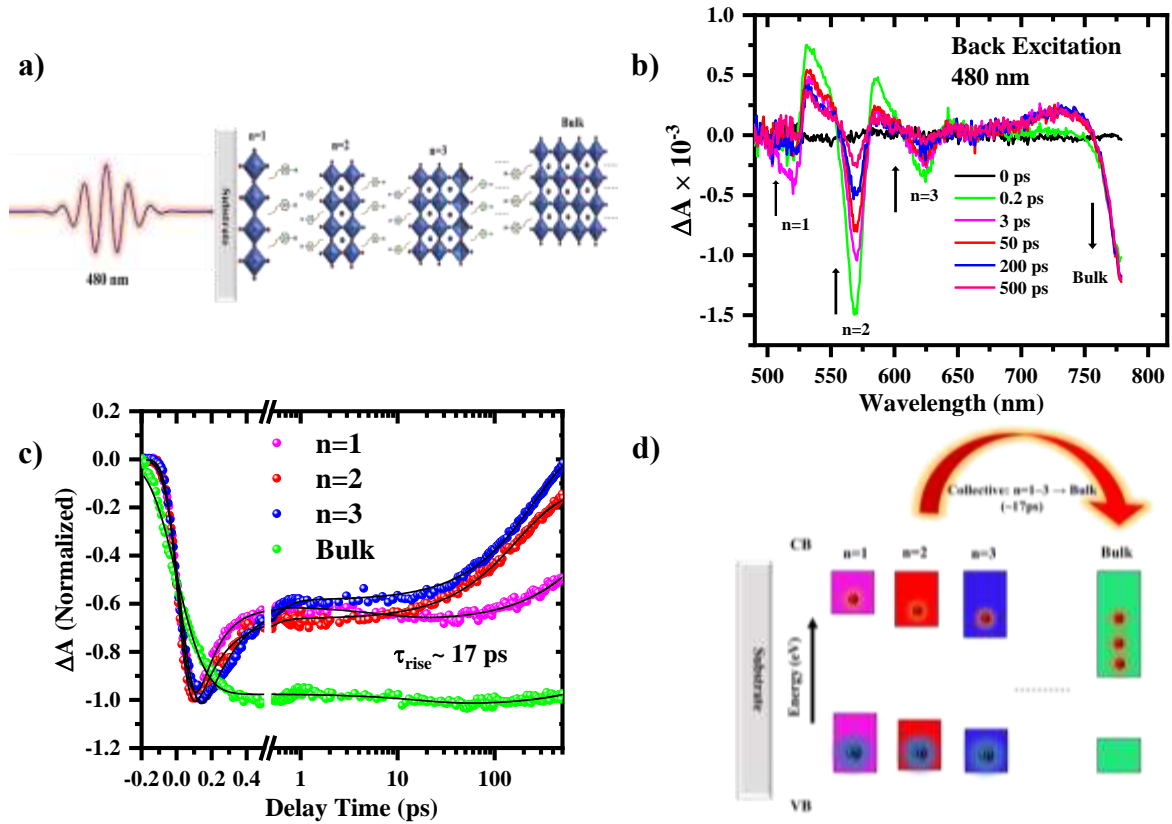

**Figure 3.** TA spectra and carrier dynamics of the 2D perovskite film under back-side excitation at 480 nm. (a) Schematic illustration of the back-excitation geometry used for TA measurements. (b) TA spectra recorded at selected pump–probe delay times, showing bleach features at 520, 570, 624, and 780 nm, corresponding to the n=1, n=2, n=3, and bulk phases, respectively. Black arrows indicate the temporal evolution of the bleach signals. (c) TA kinetics probed at the n=1, 2, 3, and bulk bands under back-excitation. The slow rising dynamics at the bulk band reflect electron accumulation in this phase via transfer from the lower-n domains. Solid lines represent exponential fits of the kinetics, yielding an electron transfer time of ~17 ps. (d) Schematic illustrating directional electron transfer from the n=1, 2, and 3 phases into the bulk phase following 480 nm excitation.

To disentangle the possibility of both carrier and energy transfer, photoluminescence excitation (PLE) spectroscopy was performed by monitoring bulk emission at 778 nm. As shown in Figure S6, the PLE spectrum (trace b) closely tracks the bulk absorption (trace a) but shows no correspondence with the discrete absorption features of small-n phases (n=1-3). This unambiguously demonstrate that excitation of low-n domains does not feed bulk emission, thereby ruling out singlet energy transfer as a viable pathway[54]. Instead, bulk emission arises solely from direct excitation of bulk-like domains. Furthermore, the type-II band alignment across the different n phases (Figure 2b) suggests the possibility of photoinduced electron transfer from the lower-n phases to the bulk phase of the 2D perovskite.

Importantly, the near-identical early-time TA decay dynamics across the n=1-3 phases (Figure 3c) argue against the conventional 'cascade transfer' picture, where carriers are expected to sequentially migrate through intermediate phases of increasing n. Instead, our data reveal a collective transfer mechanism, in which electrons originating from multiple low-n states (n=1-3) are directly injected into the bulk (Figure 3d). Given that the carriers are photoexcited at 480 nm, well above the band edge of the n=1 phase, they initially possess substantial excess energy. Upon relaxation toward the band edge, this excess energy could facilitate direct transfer from the n=1-3 states into the bulk, thereby bypassing the sequential cascade though intermediate phases. Notably, the TA kinetics of the n=1 phase (Figure 3c) exhibits a minor delayed (after 1 ps) rise component (~7 ps), which may originate from hole transfer from higher-n domains (n>1) into the n=1 phase[48].

To quantitatively validate the mechanistic framework inferred from the phase-resolved TA measurements, a compartmental kinetic model was developed using Glotaran platform[59]. In this approach, each photoactive phase n =1, 2, 3, and bulk was treated as an individual compartment, with the model explicitly accounting for both interphase population transfer and intrinsic recombination within each phase. Such a multi-channel decay framework is particularly useful for quasi-2D perovskites, where the coexistence of multiple layered domains and strongly coupled charge-transfer pathways necessitates treatment as an interconnected network of populations rather than as independent decay channels. This strategy enables direct correlation between the spectral fingerprints and the underlying carrier dynamics, providing a unified description of relaxation and transfer processes[60].

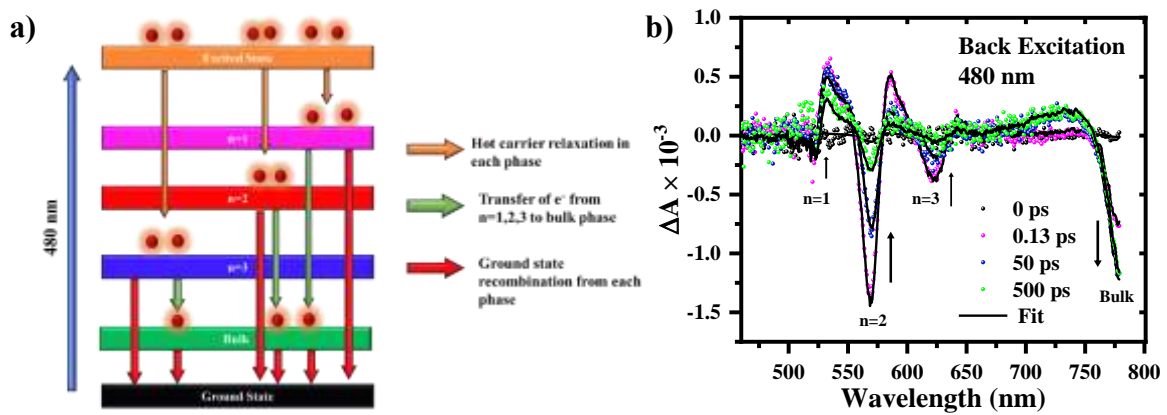

**Figure 4.** Compartmental modeling of TA spectra measured under back-side excitation at 480 nm. (a) Kinetic scheme for back-side excitation at 480 nm, where carriers relax to the band edges of small-n phases before competing between electron transfer to the bulk and intrinsic recombination. (b) Global target analysis of the corresponding TA spectra, showing that the compartmental model captures both the rapid relaxation and the branching dynamics with good agreement to the experimental data.

Under back-side excitation at 480 nm (Figure 3a), the experimental TA spectra were fitted using the kinetic model (Figure 4a) comprised four coupled compartments representing the n=1, 2, 3, and bulk phases. Following photoexcitation, carriers rapidly relax to the band edges of the small-n phases (orange arrow in Figure 4a). At this stage, two competing processes may emerge: (i) ultrafast electron transfer from the small-n phases into the bulk (green arrow), and (ii) intrinsic recombination within each small-n domain. This delicate competition dictates the carrier lifetime available for long-range transport. To capture this quantitatively, recombination pathways were explicitly incorporated into the model. Since the applied fluence of 0.3 μJ cm$^{-2}$ per pulse lies well below the threshold for nonlinear channels such as Auger recombination or exciton–exciton annihilation, these processes were excluded. Each compartment was initialized with the PB signals of its respective phases shown in Figure 3b. The time constants and branching ratios used in the modelling were obtained from fitting of TA kinetics corresponding to individual phases (see Table S1 in SI). The resulting spectral fit reproduces experimental spectra with high accuracy. Therefore, modeling of spectral data supports our picture of direct transfer of electrons from small n phases to bulk.

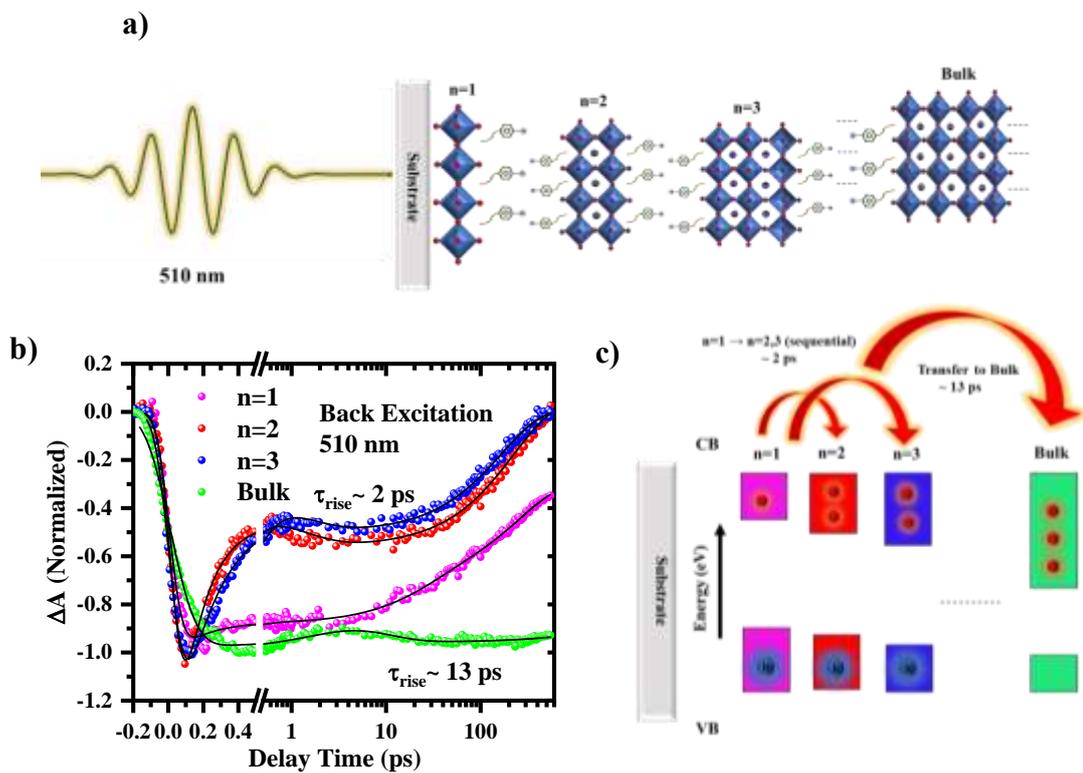

**Figure 5.** Carrier dynamics in the quasi-2D perovskite film under back-excitation at 510 nm. (a) Schematic illustration of the back-excitation configuration employed for TA measurements. (b) Kinetic traces at the bleach maxima of n=1, 2, 3, and bulk phases following selective excitation of the n=1 band-edge, showing biexponential decay in n=1,

triexponential behavior with intermediate rises in n=2 and n=3, and a prolonged rise followed by a long-lived decay in the bulk-consistent with sequential carrier transfer from n=1 to the bulk. (c) Schematic representation of the directional carrier cascade, whereby carriers excited at the n=1 band edge sequentially transfers through higher-n phases and ultimately accumulate in the bulk phase.

After investigating the case of back excitation at 480 nm (above the band-edge), we next performed TA kinetics measurement under band-edge excitation of each phase one by one to directly monitor the corresponding carrier transfer pathways. This approach enabled direct initiation of carrier dynamics within specific small-n domains and provided a phase-resolved picture of their subsequent transfer toward the bulk. Upon selective excitation using 510 nm (fluence ~ 0.342 µJ cm$^{-2}$) light corresponding to the band-edge of the n=1 phase (Figure 5a), the TA kinetics (Figure 5b) displays a triexponential decay, comprising a fast component ($\tau_1$ ~ 0.17 ps) and an intermediate and slower decay components ($\tau_2$ ~ 24 ps, $\tau_3$=260 ps, Table S2), reflecting rapid depopulation though electron transfer as well as recombination's within this phase.

In contrast, the n=2 and n=3 phases exhibit triexponential decay kinetics with a biexponential decay and an intermediate rise (Figure 5b, Table S2). Because this excitation energy is also higher than their bandgap energies, these phases are directly photoexcited, giving rise to the initial decay associated with electron transfer to the bulk. The intermediate rise (~2 ps) reflects electrons injected at the n=1 band edge, which lack sufficient excess energy to bypass intermediate phases and instead undergo sequential transfer through n=2 and n=3 before reaching the bulk (Figure 5c).

Further confirmation of sequential electron transfer has been obtained by direct band-edge excitation of the n=2 and n=3 (fluence ~ 0.27 µJ cm$^{-2}$) phases. Under 560 nm (fluence ~

0.5 µJ cm$^{-2}$) excitation, the corresponding TA kinetics (Figure S7a) displays biexponential decay in the n=2 phase (Table S3). However, the TA kinetics in n=3 and bulk phases (Figure S7a) exhibit two decays and an intermediate rise component (Table S3), consistent with a stepwise n=2→n=3→bulk transfer pathway. In contrast the selective excitation at 600 nm directly populated the n=3 phase, whose TA kinetics undergone only decay while TA kinetics of bulk phase showed an early rise followed by a decay (Figure S7b, Table S4), revealing prompt (~ 1 ps) electron transfer into bulk.

Importantly, comparison of TA kinetics of the bulk phase under different excitation wavelengths (480, 510, 560, 600, and 750 nm; Figure S8(a-d)) highlights the dual nature of the electron transfer mechanism. The high-energy excitation (480 nm) generates carriers across all small-n phases simultaneously, leading to their nearly synchronous depopulation and the appearance of cooperative transfer into the bulk. By contrast, phase-selective excitation at the band edges resolves the sequential, cascade-like handoff between adjacent phases. Finally, direct resonant excitation of the bulk band edge (750 nm) gives rise to only TA decay without delayed rises, confirming that the bulk population under above-bandgap excitation originates primarily from interphase electron transfer rather than direct pumping. Collectively these results provide a comprehensive, phase-resolved picture of electron transfer pathways in quasi-2D perovskites, establishing how the excitation conditions govern weather the electron transfer proceeds collectively or sequentially.

### 2.3.2 Under Front Excitation

Beyond the directional electron transfer from small-n domains into the bulk, the reverse process-hole migration from the bulk toward lower-n phases-also energetically feasible due to type-II band alignment. To directly probe this hole-transfer pathway, TA measurements were carried out under front-side excitation at 750 nm (fluence ~ 0.5 µJ cm$^{-2}$), where the bulk is selectively excited (Figure 6a). The corresponding TA spectra (Figure 6b) display a pronounced PB at 778 nm, characteristic of the bulk phase, accompanied by weaker bleach features at shorter wavelengths associated with the n=1, 2, and 3 phases. Among these, the n=2 and 3 phases manifest predominantly as broad PIA signatures, while the n=1 phase exhibits a distinct PB feature resembling that of the bulk. The appearance of n=2 and 3 features as PIA is consistent with earlier report[42]. With increasing delay, the bulk PB progressively decays, whereas the small-n features grow in intensity, unambiguously evidencing hole transfer from the bulk into the lower-n phases. Notably, the bleach amplitudes linked to hole transfer are considerably weaker than those observed for electron transfer, likely due to two factors: (i) the intrinsically weaker contribution of hole filling to the PB signal[61], and (ii) partial hole trapping within the bulk, which reduces the effective carrier population available for transfer[62].

The TA kinetics at the characteristic bleach positions (Figure 6c) further clarify the dynamics. The TA kinetics of the bulk phase follows a biexponential decay with time constants of 2 ps and >1 ns (Table S5). The ultrafast component corresponds to depopulation driven by hole transfer, while the slower decay reflects carrier recombination and/or trapping. In contrast, the n = 3 phase shows a sharp rise with a 0.3 ps time constant (Table S5), indicative of direct and efficient hole transfer from the bulk. The n=2 phase displays a biexponential rise, comprising a fast ~0.3 ps component (direct transfer from bulk to n=2) and a slower ~1 ns component (Table S5), consistent with phonon-assisted hole migration from the n=3 phase. Interestingly, the n=1 phase exhibits a much slower rise (~17 ps) (Table S5), suggesting that direct bulk-to-n=1 hole transfer is strongly suppressed due to the large valence-band offset[63]. Instead, holes migrate sequentially through the n=3 and n=2 phases (Figure 6d), establishing a kinetic bottleneck that delays population build up in the n=1 phase[64, 65]. Importantly, unlike earlier reports where the n=1 phase functioned primarily as a hole-blocking layer[53], our results demonstrate effective hole transfer into the n=1 phase, albeit with delayed formation kinetics. These findings highlight that interphase coupling in our system ultimately enables hole accumulation in the lowest-n phase. Moreover, the comparable timescales of electron transfer under back excitation at 480 nm and hole transfer into the n=1 phase under front excitation at 750 nm suggest a balanced spatial separation of electrons and holes, indicative of enhanced structural order and suppressed thermal disorder[66]. The role of fluorination is particularly significant in this context. With fluorine substitution at the para position of the bulky BA cation, prior theoretical studies have shown that charge transfer rates can be accelerated by one to two orders of magnitude. This enhancement arises from strengthened hydrogen bonding and dipolar interactions, which reduce reorganization energy within the Marcus regime, thereby facilitating faster electron and hole transfer[67]. Collectively, these results underscore how structural and chemical engineering of quasi-2D perovskites-particularly through fluorination-provides a powerful avenue to optimize bidirectional carrier transfer across vertically stacked phases.

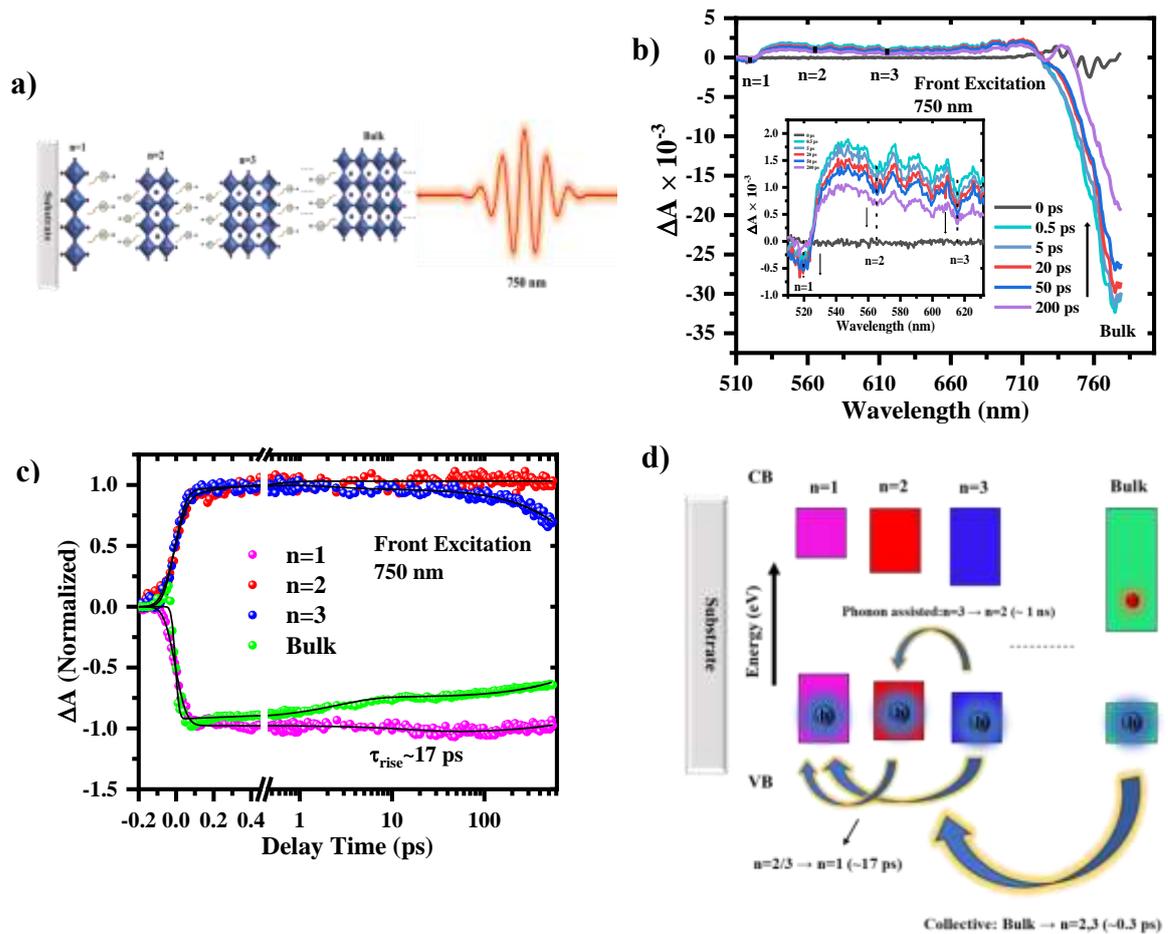

**Figure 6.** Carrier dynamics in the 2D perovskite film under front-side excitation at 750 nm. (a) Schematic illustration of the front-excitation configuration used for TA measurements. (b) TA spectra recorded at selected pump–probe delay times. The main panel shows the bleach feature at 778 nm corresponding to the bulk phase, while the inset highlights the bleach features at 520, 570, and 614 nm associated with the n=1, 2, and 3 phases, respectively. Black arrows indicate the temporal evolution of the bleach signals. (c) TA kinetics probed at the n=1, 2, 3, and bulk bleach bands under front excitation. The slow rise dynamics of the n=1, 2, and 3 bands reflect hole accumulation via transfer from the bulk, with a characteristic hole transfer time to the n=1 phase of ~17 ps. Solid lines represent exponential fits to the kinetics. (d) Schematic representation of the band alignment illustrating that, under front excitation, the type-II band offset facilitates hole transfer from the bulk to higher-n phases, eventually leading to accumulation in the n=1 phase.

Analogous to the back-excitation case at 480 nm, we also developed a compartmental kinetic model under front-side excitation at 750 nm, where the kinetic framework was inverted to describe hole transfer from the bulk into the layered phases (Figure 7a). In this case, carriers are generated exclusively in the bulk and subsequently evolve via two pathways: (i) interphase hole transfer into lower-n domains (pink arrow), and (ii) intrinsic recombination within the bulk (red arrow). The observed TA kinetics (Figure 6c) indicate a sequential migration of holes, first into the n=3 compartment, then into n=2, and finally into n=1. To reproduce the biexponential rise observed for the n=2 phase, an additional phonon-assisted transfer channel (n=3→n=2, blue arrow) was included. The associated time constants and branching ratios of each of these processes were obtained from the single-phase kinetic fitting of each phase under 750 nm excitation (see Table S5 in SI). The resulting spectral fits (Figure 7b) reproduce the experimental spectra with high accuracy. Therefore, modeling of the spectral data supports our proposition of stepwise hole transfer from bulk to the small-n phases.

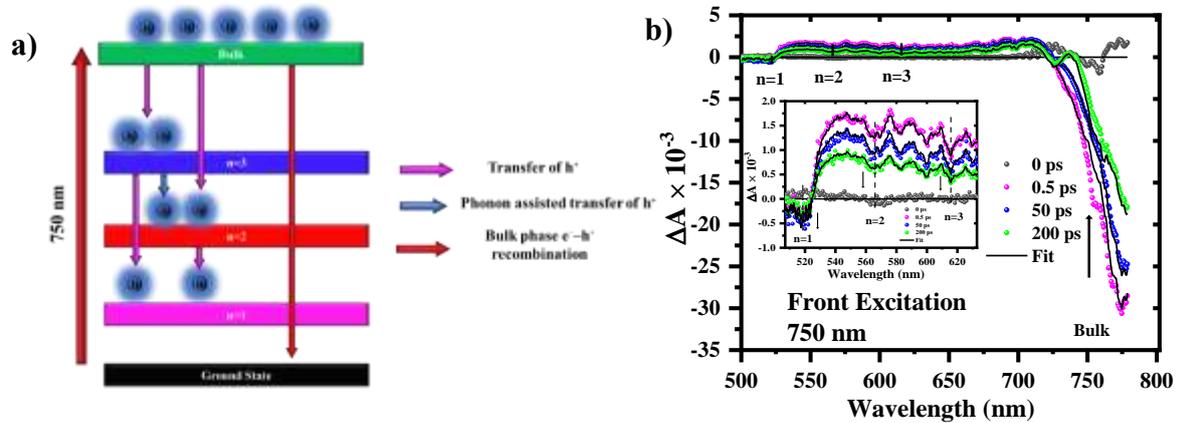

**Figure 7.** Compartmental modeling of TA spectra under front-side excitation. (a) Kinetic framework for front-side excitation at 750 nm, where photoexcited carriers originate in the bulk and undergo sequential hole transfer through n=3 and n=2 before eventually accumulating in the n=1 phase. Intrinsic recombination in the bulk and phonon-assisted transfer channels (e.g., n=3→n=2) were explicitly included to reproduce the observed biexponential features. (b) Global target analysis of the TA spectra under front excitation, validating the sequential hole migration pathway and revealing the delayed ~17 ps buildup of the n=1 bleach as evidence of a kinetic bottleneck imposed by stepwise transfer.

Taken together, the integration of phase-resolved spectroscopy with compartmental modelling demonstrates how quasi-2D perovskites can be engineered to promote cooperative electron–hole pathways, thereby enabling enhanced spatial charge separation across vertically stacked phases.

## 2.3.4 Implications for Photocurrent Generation

To explore the practical implications of bidirectional carrier transfer revealed by the TA study, diode devices were fabricated using the perovskite, $(4\text{-FBA})_2(MA)_{n-1}Pb_nI_{3n+1}$. Both electron-only and hole-only devices were fabricated (see Note S5, Device Fabrication in the SI). The electron-only device was constructed with the architecture ITO/SnO$_2$ layer functions as an efficient hole-blocking layer, while the Ag electrode serves as the electron-collecting contact, thus allowing primarily electron transport. Under forward bias (positive bias on the Ag electrode), the device exhibits a dark current of $5.89 \times 10^{-6}$ A at 50 mV (Figure 8a). Upon illumination from the ITO side-corresponding in the TA measurements-the current increased to $1.36 \times 10^{-5}$ A (at 50 mV) under 750 nm excitation, and further rose significantly to $5 \times 10^{-5}$ A (at 50 mV) under 480 nm excitation. The efficient photocurrent generation can be rationalized by the transfer of photogenerated electrons through different layers of the 2D perovskite film. Backside illumination at 750 nm primarily excites carriers in the bulk phase. The photogenerated electrons then migrate downhill toward the Ag electrode, producing a photocurrent that nearly doubles the total current. In contrast, 480 nm excitation generates electrons in the small-n phases, which subsequently transfer downhill in to the bulk before being collected at the Ag contact, leading to an order-of-magnitude enhancement in the photocurrent (Figure S9a). The higher photocurrent observed under 480 nm illumination compared to 750 nm arises because the shorter-wavelength excitation generates carriers across all perovskite phases, whereas the longer wavelengths primarily excite the bulk phase.

Similarly, a hole-only device was fabricated with the structure ITO/PEDOT:PSS/perovskite/Spiro-OMeTAD/Ag (Figure S9b). In this configuration, PEDOT:PSS serves as a hole-extracting and transporting layer, with ITO acting as the hole-collecting electrode, while while Spiro-OMeTAD functions as an efficient electron-blocking layer.

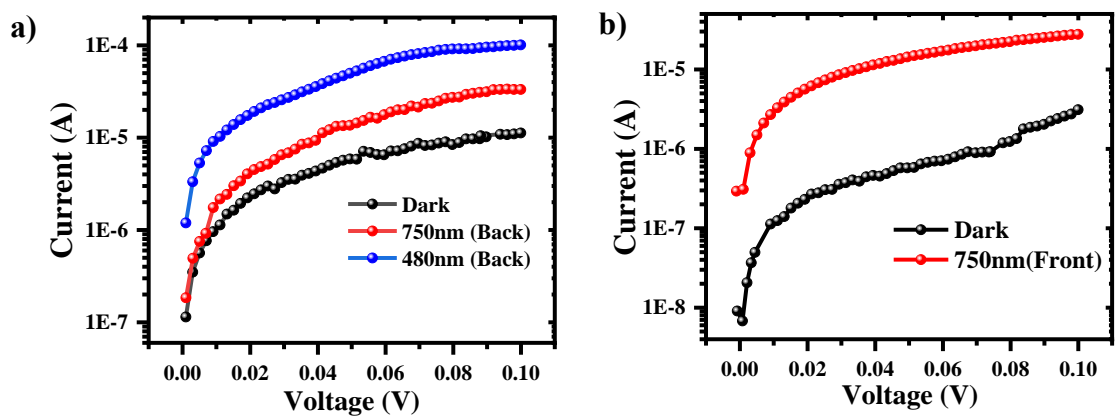

**Figure 8.** Current–voltage (I–V) characteristics of (a) the electron-only device measured under dark conditions and illumination at 750 nm and 480 nm from the ITO side (back excitation), and (b) the hole-only device measured under dark conditions and illumination at 750 nm from the Spiro-OMeTAD side (front excitation).

Upon illumination at 750 nm from the Spiro-OMeTAD side-corresponding to front-side excitation in the TA measurements-the device exhibited a pronounced photocurrent of $1.44 \times 10^{-5}$ A at 50 mV, compared to a dark current of $5.76 \times 10^{-7}$ A (Figure 8b). In the hole-only device, under 750 nm front-side illumination, the light is predominantly absorbed by the bulk phase, generating photocarriers. The photogenerated holes migrate energetically downhill toward n=1 layers and are subsequently extracted through the PEDOT:PSS layer to the ITO electrode. This transport is facilitated by the favorable alignment of the VBM across the different n phases with the highest occupied molecular orbital (HOMO) level of the PEDOT: PSS, while electron flow toward the Ag electrode is effectively suppressed (Figure S9b).

Taken together, these results demonstrate that the bidirectional carrier transfer observed in ultrafast TA measurements is faithfully reproduced at the device level through single-carrier conduction. Furthermore, the transfer of photogenerated electrons and holes in opposite directions-driven by exciton dissociation at the interfaces of different n layers-results in efficient photocurrent generation. The strong correlation between ultrafast transfer dynamics and steady-state device behaviour underscores the critical role of interfacial phase engineering in optimizing charge extraction and enhancing the performance of layered perovskite-based optoelectronic devices.

## 3. Conclusions

In summary, transient absorption spectroscopy and kinetic modeling reveal bidirectional photoinduced carrier transfer in fluorinated quasi-2D Ruddlesden-Popper perovskites, with electrons flowing from small-n phases to the bulk under back-side excitation and holes migrating from the bulk to lower-n phases under front-side excitation. Strong interphase coupling enables hole population even in the n=1 phase, overcoming previously assumed bottlenecks. Electron-only and hole-only devices confirm that these directional pathways translate directly to efficient carrier extraction, demonstrating effective spatial separation of electrons and holes between the n=1 and the bulk phase. These findings underscore the importance of phase engineering and interphase interactions in quasi-2D perovskites and provide a strategy for designing next-generation optoelectronic devices with optimized charge transport.

## 4. Acknowledgements

The authors gratefully acknowledge the Advanced Materials Research Centre (AMRC) at the Indian Institute of Technology Mandi for providing the necessary experimental facilities. S.H. and K.G. acknowledge fellowship support the Ministry of Education (MoE), and N.S.K. acknowledges the Department of Science and Technology (DST), Government of India, for fellowship support under the INSPIRE program. S.K.P. is thankful to the Science and Engineering Research Board (SERB), Government of India, for financial support (Grant No. CRG/2022/006320). The authors also acknowledge Dr. Prakriti Ranjan Bangal at the CSIR-Indian Institute of Chemical Technology (CSIR-IICT) for his assistance with compartmental modeling. We further thank ChatGPT (Open AI) for its help in improving the English language of this manuscript.

# Bidirectional Photoinduced Carrier Transfer in Fluorinated Quasi-2D Perovskites Governing Enhanced Photocurrent Generation

*Soumya Halder [1,2][†], Koushik Gayen [1,2][†], Nagendra S. Kamath [1,2], and Suman Kalyan Pal [1,2]\**

[1]*School of Physical Sciences, Indian Institute of Technology Mandi, Kamand, Mandi-175005, Himachal Pradesh, India*

[2]*Advanced Materials Research Centre, Indian Institute of Technology Mandi, Kamand, Mandi-175005, Himachal Pradesh, India*

[†]These authors contributed equally to this work.

AUTHOR INFORMATION

**Corresponding Author**

*E-mail: suman@iitmandi.ac.in; Phone: +91 1905267040

# Note S1: Experimental Methods

## Materials

4-Fluorobenzylamine (4-FBA, 98%), methylammonium iodide (MAI, 99.9%), and lead(II) iodide (PbI$_2$, 99.99%) were obtained from TCI Pvt. Ltd. Solvents including dimethylformamide (DMF), dimethyl sulfoxide (DMSO), hydroiodic acid (HI, 57%), anhydrous ethanol, isopropyl alcohol, chlorobenzene, acetonitrile, toluene, acetone, and deionized water were used as received. Tin(IV) oxide (SnO$_2$), poly(3,4-ethylenedioxythiophene):polystyrene sulfonate (PEDOT:PSS), bis(trifluoromethane)sulfonimide lithium salt (Li-TFSI), 4-tert-butylpyridine (TBP), the cobalt complex tris(2-(1H-pyrazol-1-yl)-4-tert-butylpyridine)cobalt(III) tris(bis(trifluoromethylsulfonyl)imide) (FK209), and the hole-transport material 2,2′,7,7′-tetrakis(N,N-di-p-methoxyphenylamine)-9,9′-spirobifluorene (Spiro-OMeTAD) were purchased from Sigma-Aldrich. All reagents and solvents were used without further purification.

## Synthesis of 4-Fluorobenzylamine Iodide (4-FBAI)

To synthesize 4-fluorobenzylamine iodide (4-FBAI), 4-fluorobenzylamine (4-FBA) was reacted with an equimolar amount of hydroiodic acid (HI) in ethanol. The reaction was conducted under continuous stirring for 2 hours while maintaining the temperature in an ice bath to minimize side reactions and improve product yield. Upon completion, the solvent was removed using rotary evaporation, yielding the crude intermediate. The obtained solid was then redissolved in a minimal amount of ethanol and washed several times with diethyl ether to remove unreacted precursors and by-products. Finally, the purified white crystalline 4-FBAI was dried thoroughly and stored under a nitrogen atmosphere for subsequent use.

## Preparation of Quasi-2D Perovskite (4-FBA)$_2$(MA)$_{n-1}$Pb$_n$I$_{3n+1}$ (n=3) Precursor Solution

The (4-FBA)$_2$MA$_{n-1}$Pb$_n$I$_{3n+1}$ (n=3) perovskite precursor solution was prepared by dissolving stoichiometric amounts of 4-fluorobenzylammonium iodide (4-FBAI), methylammonium iodide (MAI), and lead iodide (PbI$_2$) in anhydrous DMF/DMSO (9:1, v/v) at a molar ratio of 2:2:3[1]. The mixture was stirred at 50 °C for 12 hours to ensure complete dissolution and obtain a homogeneous and clear solution. After cooling to room temperature, the solution was filtered through a 0.45 μm PTFE syringe filter to remove any residual particulates. The filtered precursor was then stored in sealed vials under an inert atmosphere inside a glovebox to prevent degradation from moisture and light exposure. The resulting stable and transparent precursor solutions were subsequently used for thin-film deposition.

## Perovskite Film Preparation

Quartz substrates were sequentially cleaned by ultrasonication in a series of solvents: first in an aqueous soap solution, followed by isopropyl alcohol (IPA) and acetone, each for 10 minutes. The cleaned substrates were then treated with ultraviolet/ozone (UV/O$_3$) for 20 minutes to remove residual organic contaminants and enhance surface wettability. Subsequently, the substrates were annealed at 130 °C for 10 minutes to ensure complete drying.

The perovskite precursor solution was then spin-coated onto the preheated substrates at 4000 rpm for 20 seconds to form uniform thin films. After deposition, the films were annealed at 100 °C for 5 minutes to facilitate crystallization and ensure complete solvent evaporation.

## Characterization

X-ray diffraction (XRD) patterns of the $(4\text{-}FBA)_2(MA)_{n-1}Pb_nI_{3n+1}$ (n = 3) perovskite films were recorded using a Rigaku MiniFlex diffractometer. Steady-state UV–visible absorption spectra were obtained with a Shimadzu UV-2700 spectrophotometer. Photoluminescence (PL) measurements were performed using a home-built micro-PL setup, where a 5X, 0.15 NA microscope objective was used to collect the emitted light. The spectra were recorded using a charge-coupled device (CCD) detector equipped with a 600 grooves $mm^{-1}$ grating (Ocean Optics USB4000).

Surface morphology and microstructural analysis of the perovskite films were conducted using a NanoSEM 450 field-emission scanning electron microscope (FE-SEM), which provides ultra-high imaging resolution without the specimen size restrictions typical of conventional in-lens FE-SEMs due to its advanced electron optics design. Atomic force microscopy (AFM) measurements were carried out using a Park Systems NX7 to investigate the surface topography and roughness of the films.

Furthermore, ultraviolet photoelectron spectroscopy (UPS) measurements were performed using a Thermo Scientific NEXSA Surface Analysis System to determine the band alignment between different phases of the perovskite samples.

Current-voltage (I-V) measurements of fabricated devices were performed using a programmable source meter (Keithley 2400).

## Femtosecond Transient Absorption Spectroscopy

The femtosecond broadband transient absorption (TA) spectroscopy system was based on a Ti:sapphire amplifier (Spitfire Ace, Spectra Physics) delivering pulses with a central wavelength of ~800 nm and a pulse width of <35 fs[2, 3]. The amplifier output was split into two beams to generate the pump and probe pulses.

The pump pulses (at 480, 510, 570, 610 and 750 nm) were produced using a nonlinear optical parametric amplifier (TOPAS from Light Conversion). A small fraction of the 800 nm fundamental beam was focused onto a sapphire crystal to generate a white-light continuum (WLC), which served as the probe beam from TA spectra measurements. To ensure a stable and continuous WLC, the intensity and beam size of the 800 nm light were precisely controlled using an iris diaphragm and neutral density filters.

The probe beam was then split into two parts-a sample beam and a reference beam-which were detected simultaneously to minimize common-mode noise. A mechanical chopper operating at 500 Hz modulated the pump beam, allowing differential detection between pump-on and pump-off conditions. The time delay between the pump and probe pulses was controlled using a motorized optical delay stage.

The transmitted probe light was dispersed by a grating spectrograph (Acton SpectraPro SP 2358) and detected by a CCD array. The group velocity dispersion (GVD) in the WLC probe was corrected using a chirp correction program (Pascher Instruments). For TA kinetics measurements, monochromatic probe pulses from an additional TOPAS unit were used, and the resulting transient signals were recorded using two matched photodiodes with variable gain.

**Note S2: Determination of Pump Fluence**

The energy fluence ($F_{\text{pump}}$) of the laser beam incident on the sample was calculated using the following expression

$$F_{\text{pump}} \ (\text{J/cm}^2) = \frac{X \times 10^{-6}}{R \times A} \tag{S1}$$

where X is the average laser power at the sample position (in μW), R is the half of the repetition rate of the laser (in Hz), and A is the cross-sectional area of the pump beam (in cm²). The pump fluence was finally expressed in μJ/cm² units for convenience.

**Note S3: Analysis of Electron and Hole Transfer Mechanisms**

To elucidate the charge transfer pathways among the different phases, TA kinetic traces were analyzed by fitting the experimental data with a multi-exponential function expressed as[5, 6]:

$$\Delta A(t) = \sum_{i=1}^{n} A_i e^{-t/\tau_i} \quad (S2)$$

where $A_i$ and $\tau_i$ represents the amplitude and time constant of the $i^{th}$ component, respectively and n is total number of components. The sign of the amplitude ($A_i$) indicates the nature of the process: a positive $A_i$ corresponds to a rising component (population increase), whereas a negative $A_i$ indicates a decaying component (population decreases).

For Electron transfer analysis under 480 nm back-side excitation, the small phases exhibit a biexponential decay (Table S1), where the fast component ($\tau_1$) corresponds electron transfer from small-n phases, and a slow component ($\tau_2$) represents carrier recombination within each phase. In contrast, the bulk phase shows a rising component associated with $\tau_1$, indicating the arrival of transferred electrons from the small-n phases, followed by a long decay component ($\tau_2$) attributed to recombination of carriers in the bulk.

Under band edge excitations such as 510 nm, the n=1 phase exhibits a biexponential decay characterized by $\tau_1$ and $\tau_2$, while the n=2 and n=3 phases display an initial decay ($\tau_1$), an intermediate rise ($\tau_2$) and a long decay component ($\tau_3$). This behaviour shows the sequential transfer of electrons through n=2 and n=3 to bulk. Similar trends are observed for other band-edge excitations at 560 nm and 600 nm (Tables S3 and S4), confirming consistent electron transfer dynamics across different excitation conditions.

To analyze the hole transfer mechanism, TA kinetics measured under 750 nm front excitation were fitted using equation S6. As evident from the fitting results (Table S5), the bulk phase exhibits a biexponential decay with a fast component ($\tau_1$) and a slower component ($\tau_2$), indicating rapid depopulation followed by carrier recombination. Conversely, the n=3, n=2 and n=1 phases exhibit a rising component ($\tau_1$), signifying hole transfer from the bulk phase toward smaller n phases, followed by a long decay ($\tau_2$) for n=1 and n=3 phase attributed to carrier recombination and rise for n=2 which is attributed to phonon assisted hole transfer from n=3 to n=2. This fitting-based analysis thus enables clear identification of the directionality and timescales of both electron and hole transfer processes across the different phases by correlating the sign and magnitude of $A_i$ with the physical nature of the underlying transitions.

## Note S4: Global Analysis using Glotaran:

The transient absorption data were globally analyzed using Glotaran, a Java-based graphical user interface for the TIMP (Time-resolved Imaging and Modeling Package) platform[7]. In this approach, a compartmental model was employed to describe the kinetics of the transient species. Each transient species is represented as an individual compartment, and the transitions between these compartments are characterized by specific kinetic rate constants, which correspond to the off-diagonal elements of the transfer matrix **K.** Each column in the transfer matrix represents a particular compartment, and the inverse of the eigenvalues of the **K** matrix corresponds to the lifetimes of the respective compartments.

A linear compartmental model with *n* components can be mathematically described by the following differential equation[7]

$$\frac{d\boldsymbol{C}(t)}{dt} = \boldsymbol{KC}(t) + i(t) \qquad (S3)$$

where $\boldsymbol{C}(t) = [C_1(t), C_2(t), C_3(t), ..., C_n(t)]^T$ represents the concentration vector of all compartments, and $i(t)$ is the Instrument Response Function (IRF). The solution of this equation involves an exponential function convoluted with the IRF. Accurate definition of the **K** matrix, scaling parameters, and input compartments is essential for reliable fitting within the compartmental model framework.

## Note S5 Device Fabrication

### Electron only device

Indium tin oxide (ITO) substrates were sequentially cleaned by ultrasonication in a soap solution, deionized (DI) water, acetone, and isopropyl alcohol (IPA) for 15 min each. The cleaned substrates were then treated with ultraviolet-ozone (UV-$O_3$) for 20 min to remove residual organic contaminants and enhance surface wettability. A compact tin (IV) oxide ($SnO_2$) electron transport layer was deposited by spin-coating at 3000 rpm for 20 s, followed by annealing at 150 °C for 25 min. The (4-FBA)$_2$(MA)$_2$Pb$_3$I$_{10}$ perovskite precursor solution was subsequently spin-coated onto the preheated $SnO_2$-coated substrates at 5000 rpm for 30 s. Finally, a silver (Ag) electrode (thichness~80 nm) was thermally evaporated under high vacuum ($< 5 \times 10^{-6}$ Torr) through a shadow mask to complete the fabrication of electron-only device having structure, ITO/$SnO_2$/(4FBA)$_2$(MA)$_{n-1}$Pb$_n$I$_{3n+1}$/Ag.

### Hole only device

ITO substrates were treated with UV-$O_3$ for 20 min to remove organic contaminants and render the surface hydrophilic, facilitating uniform spreading of the aqueous PEDOT: PSS layer. The PEDOT:PSS solution was filtered through a 0.45 µm syringe filter prior to deposition and spin-coated onto the substrates, followed by thermal annealing at 140 °C for 20 min to ensure complete solvent removal and improve polymer chain ordering.

Subsequently, the (4-FBA)$_2$MA$_2$Pb$_3$I$_{10}$ perovskite precursor solution was spin-coated onto the preheated PEDOT: PSS layer at 5000 rpm for 30 s. A hole transport layer (HTL) was then deposited by spin-coating a spiro-OMeTAD solution (73 mg mL$^{-1}$ in chlorobenzene) doped with Li-TFSI (28 µL; 530 mg mL$^{-1}$ in acetonitrile), TBP (28 µL), and FK209 (18 µL; 300 mg mL$^{-1}$ in acetonitrile) at 3500 rpm for 30 s. Finally, a silver (Ag) electrode (thichness~80 nm) was thermally evaporated under high vacuum (< 5 × 10$^{-6}$ Torr) through a shadow mask, completing the hole-only device (ITO/PEDOT:PSS/(4-FBA)$_2$(MA)$_{n-1}$Pb$_n$I$_{3n+1}$/Ag) fabrication.

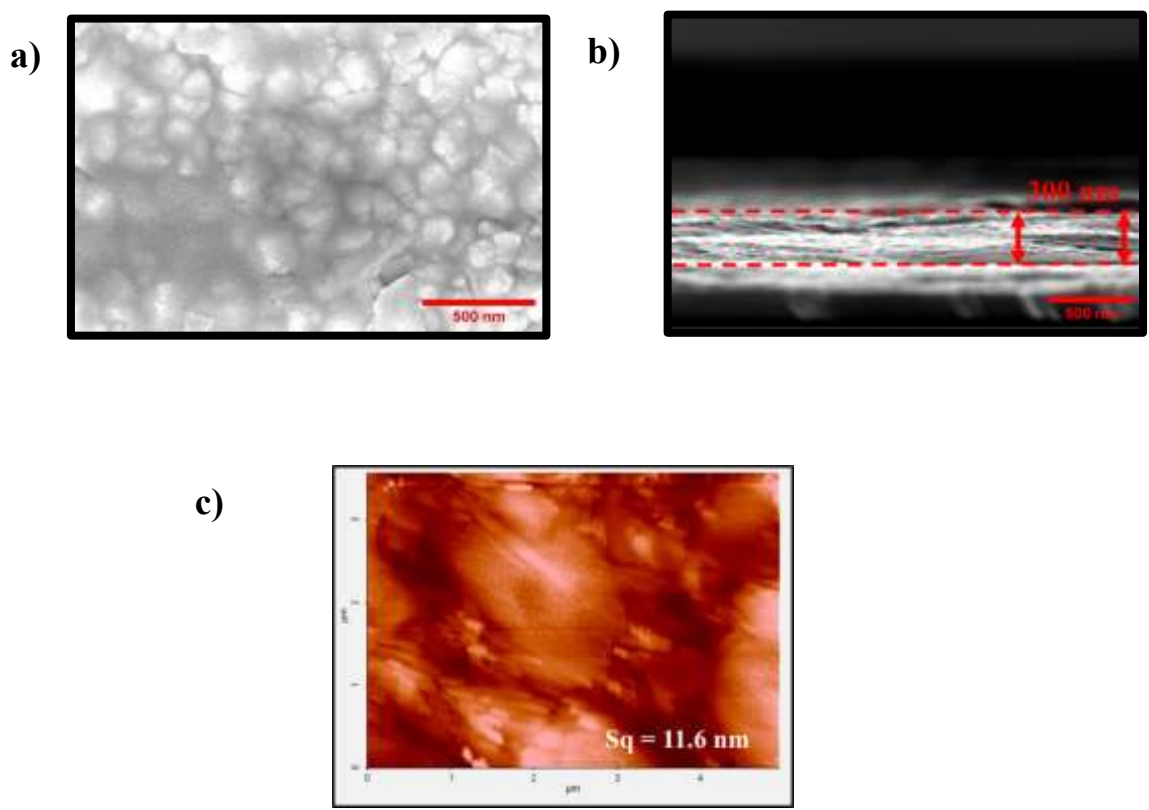

**Figure S1**. Morphological characterization of (4-FBA)$_2$(MA)$_{n-1}$Pb$_n$I$_{3n+1}$ (n=3) perovskite thin films: (a) top-view of scanning electron microscopy (SEM) image showing the surface morphology, (b) atomic force microscopy (AFM) image indicating a root-mean-square (Sq) roughness of 11.6 nm, and (c) cross-sectional SEM image used to determine the film thickness, which is approximately 300 nm.

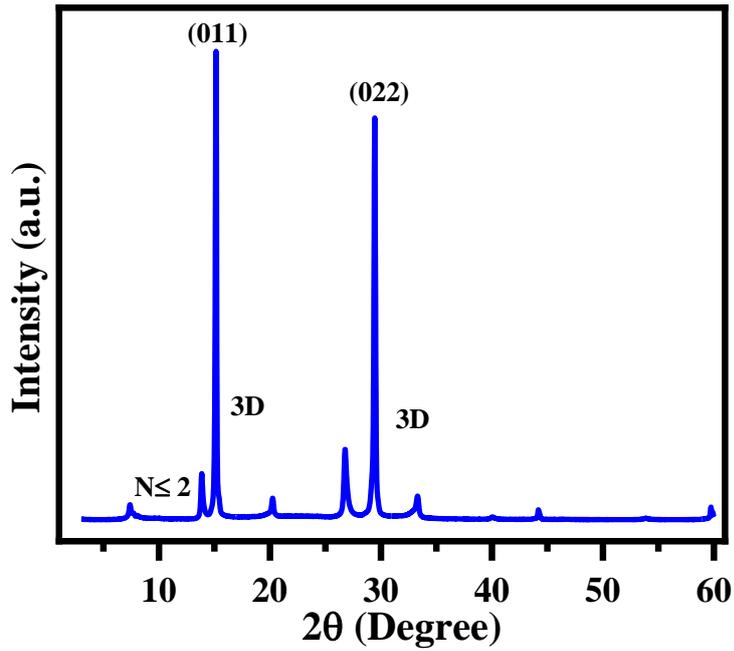

**Figure S2**. XRD analysis of the crystallographic structure of the 2D perovskite film reveals prominent peaks at low angles, confirming the formation of layered 2D perovskite phases.

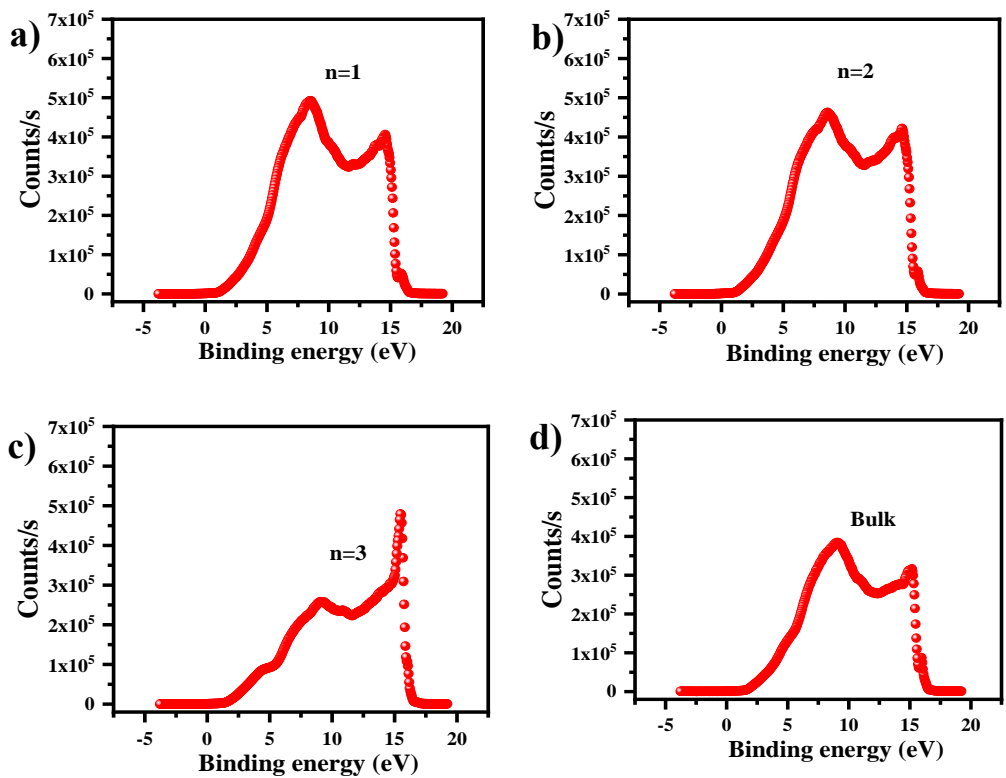

**Figure S3.** Ultraviolet photoelectron spectroscopy (UPS) spectra of $(4\text{-FBA})_2(MA)_{n-1}Pb_nI_{3n+1}$ films for (a) n = 1, (b) n = 2, (c) n = 3, and (d) the bulk phase.

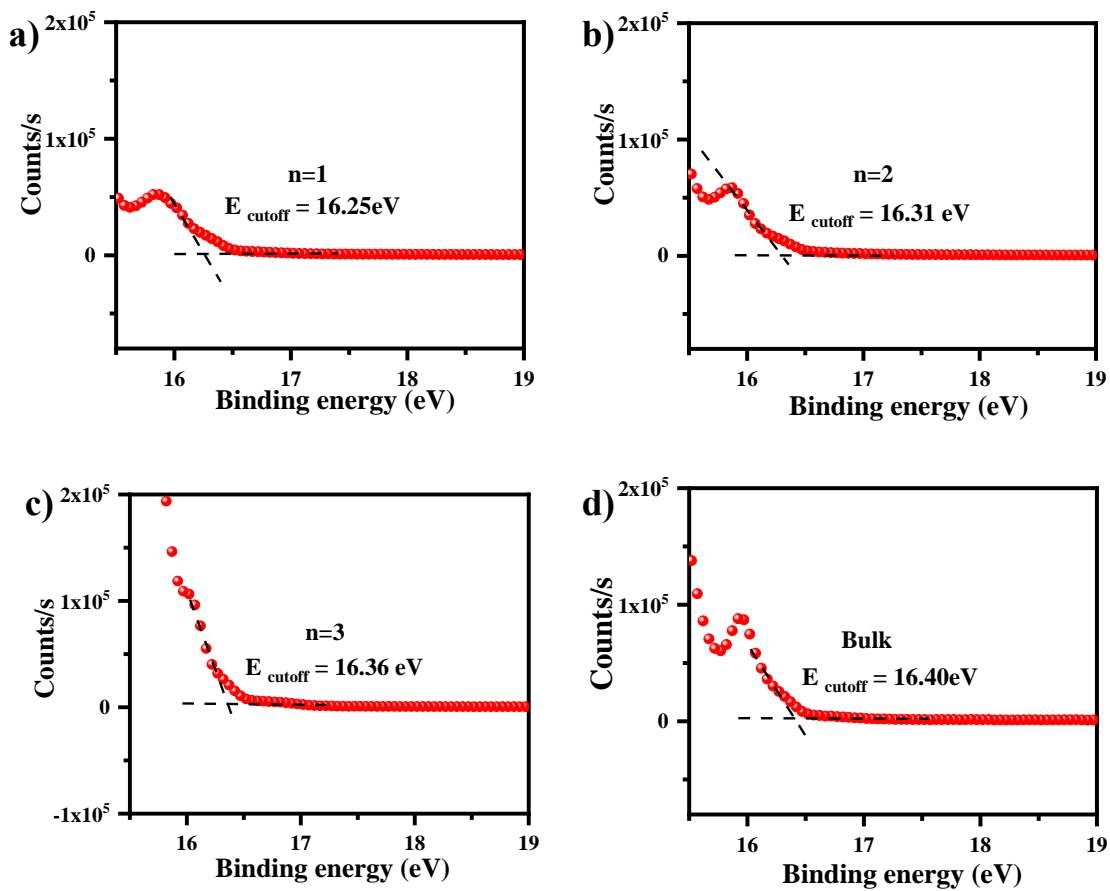

**Figure S4.** Enlarged view of the secondary electron cutoff region from the UPS spectra for (a) n = 1, (b) n = 2, (c) n = 3, and (d) the bulk phase of $(4\text{-FBA})_2(MA)_{n-1}Pb_nI_{3n+1}$ perovskite films.

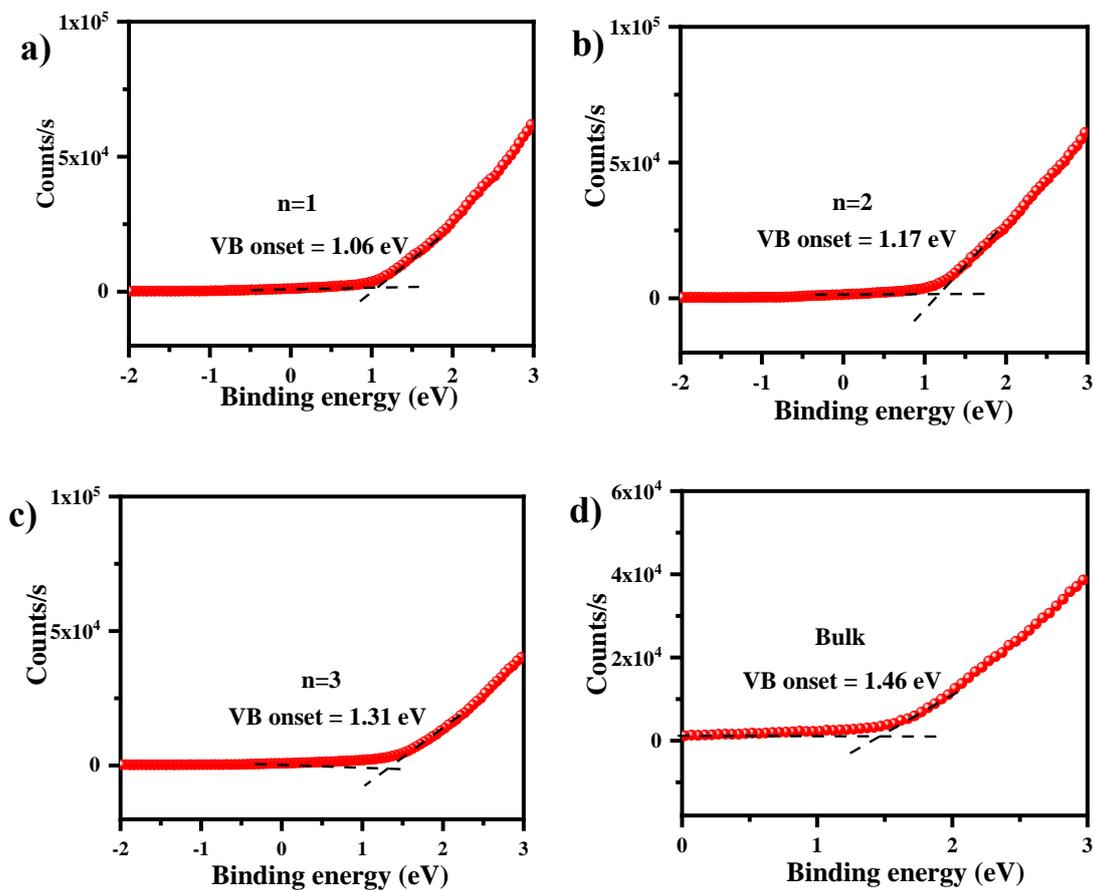

**Figure S5.** Enlarged view of the valence band onset region from the UPS spectra for (a) n = 1, (b) n = 2, (c) n = 3, and (d) the bulk phase of $(4\text{-}FBA)_2(MA)_{n-1}Pb_nI_{3n+1}$ perovskite films.

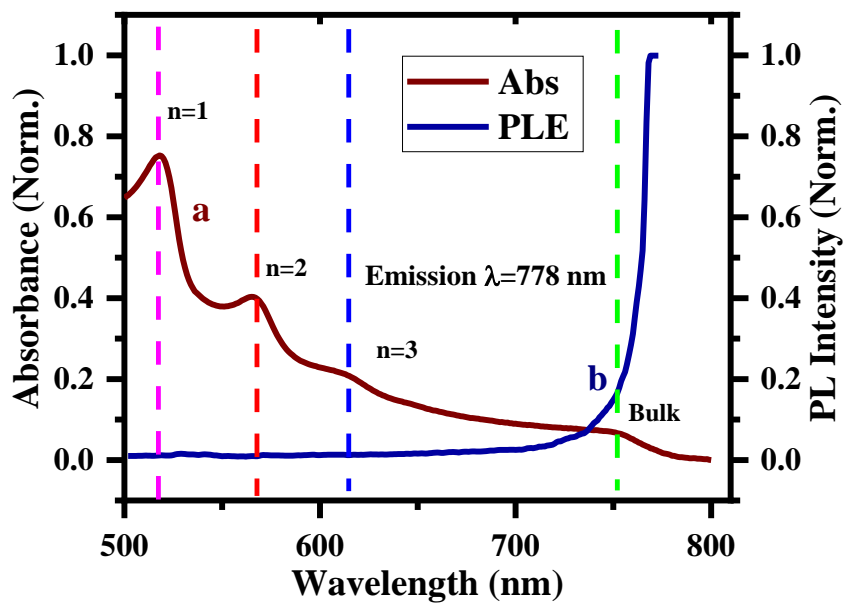

**Figure S6** Absorption (trace a) and photoluminescence excitation (PLE, trace b) spectra of the quasi-2D perovskite film. PLE spectroscopy was performed by monitoring bulk

emission at 778 nm while varying the excitation wavelength across 500–765 nm. The PLE spectrum follows the bulk absorption but shows no correspondence with the discrete excitonic features of small-n phases (n = 1–3), confirming that bulk emission originates from direct excitation of bulk-like domains rather than energy transfer from low-n phases.

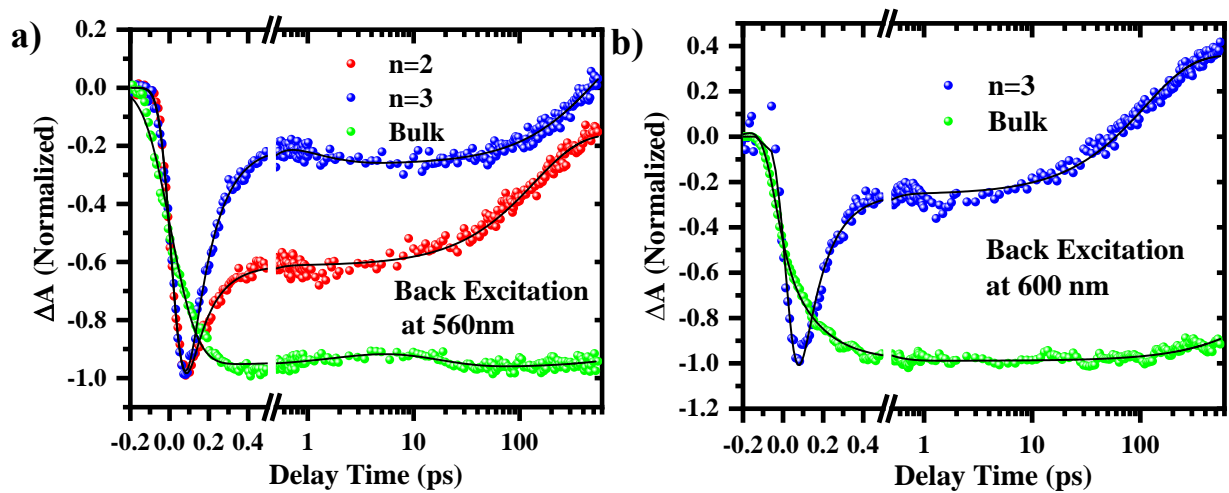

**Figure S7.** Normalized TA kinetics of quasi-2D perovskite thin films under back-side excitation at (a) 560 nm and (b) 600 nm, corresponding to the band edges of n = 2 and n = 3 phases, respectively.

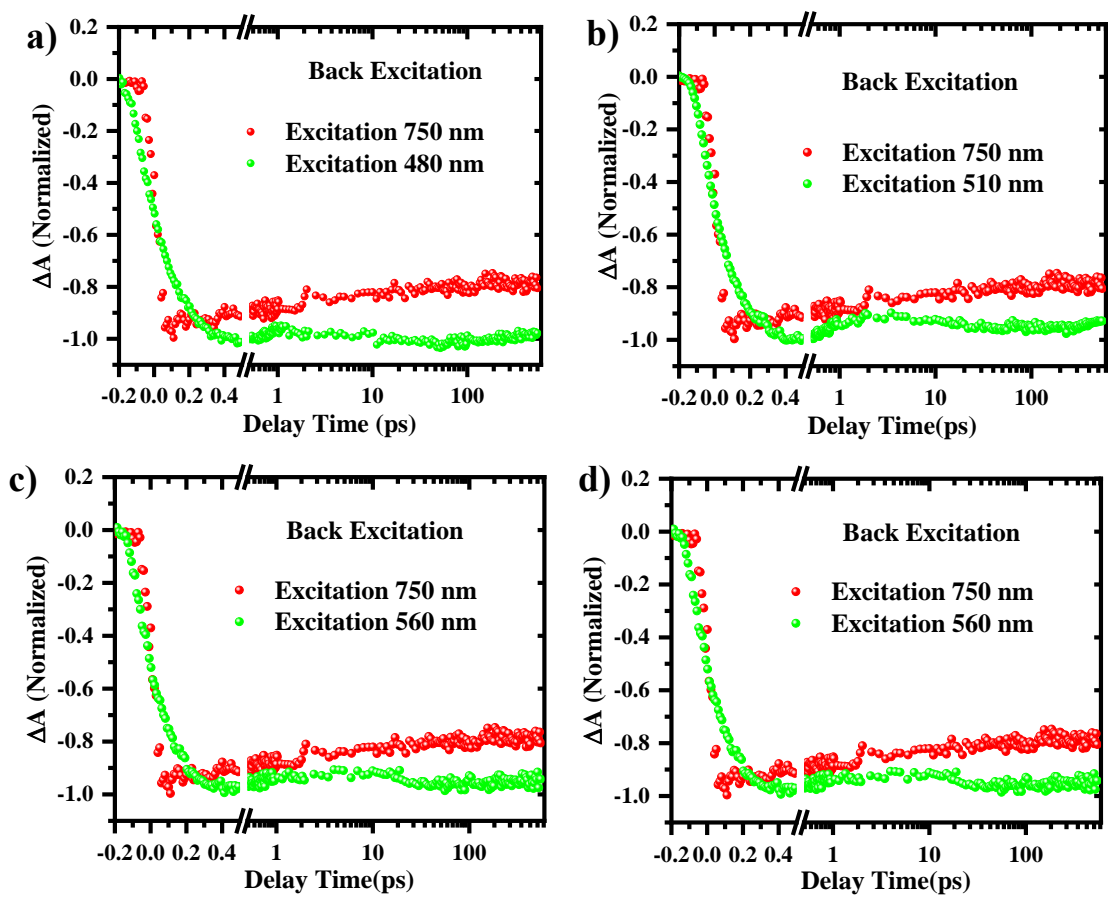

**Figure S8.** Comparison of bulk-phase TA kinetics under different excitation conditions: (a) 480 and 750 nm, (b) 510 and 750 nm, (c) 560 and 750 nm, and (d) 600 and 750 nm. The distinct kinetic behaviors reveal the dual nature of the photoinduced electron transfer mechanism in the quasi-2D perovskite film.

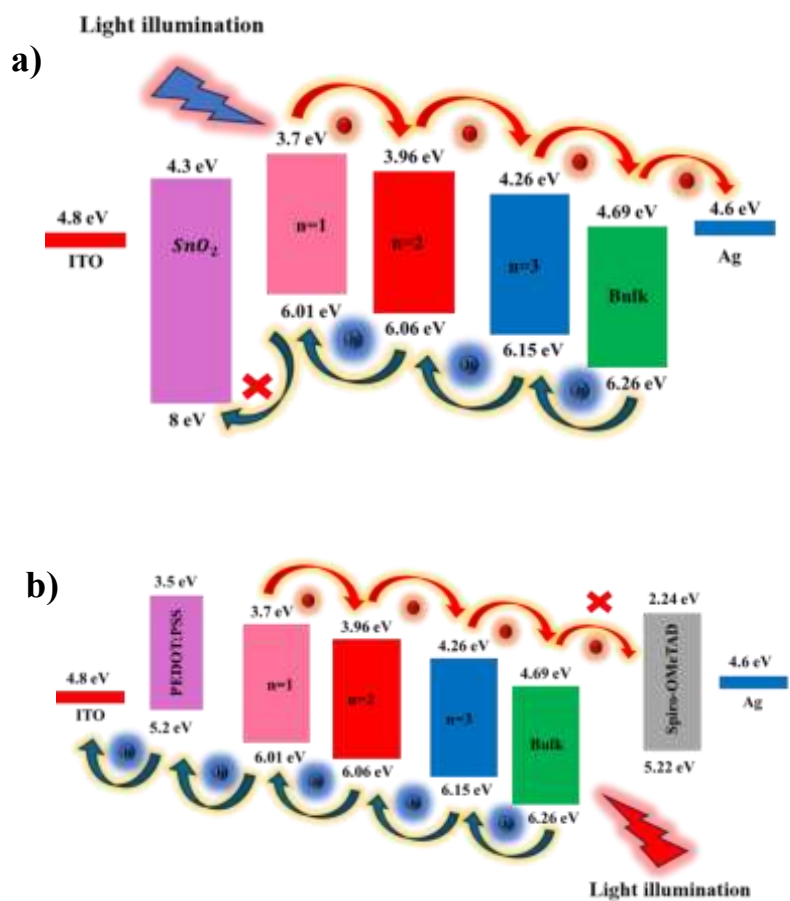

**Figure S9.** Schematic representation of the mechanism of photocurrent generation in (a) the electron-only device (ITO/SnO$_2$/perovskite/Ag) and (b) the hole-only device (ITO/PEDOT:PSS/perovskite/Spiro-OMeTAD/Ag.

Table S1. Fitting parameters for the TA kinetics shown in Figure 2c measured under back-side excitation at 480 nm. The decay dynamics were fitted using a multi-exponential function described by equation (S2).

| Phase | $\tau_1$ (ps) | $A_1$ (%) | $\tau_2$ (ps) | $A_2$ (%) | $\tau_3$ (ps) | $A_3$ (%) |
|---|---|---|---|---|---|---|
| n=1 | 0.2 | 70 (Decay) | 7 | 3 (Rise) | 1000 | 27 (Decay) |
| n=2 | 0.17 | 61 (Decay) | 264 | 39 (Decay) | — | — |
| n=3 | 0.2 | 61 (Decay) | 175 | 39 (Decay) | — | — |
| Bulk | 17 | 17 (Rise) | >1000 | 83 (Decay) | — | — |

**Table S2.** Fitting parameters for the TA kinetics shown in Figure 4b (obtained under back-side excitation at 510 nm). The decay dynamics were fitted using the multi-exponential function described by equation (S2).

| Phase | $\tau_1$ (ps) | $A_1$(%) | $\tau_2$ (ps) | $A_2$(%) | $\tau_3$ (ps) | $A_3$(%) |
|---|---|---|---|---|---|---|
| n=1 | 0.17 | 21 (Decay) | 24 | 23 (Decay) | 260 | 56 (Decay) |
| n=2 | 0.15 | 65 (Decay) | 2 | 7 (Rise) | 223 | 18 (Decay) |
| n=3 | 0.24 | 65 (Decay) | 2 | 8 (Rise) | 166 | 27 (Decay) |
| Bulk | 2 | 36 (Decay) | 13 | 46 (Rise) | 2000 | 18 (Decay) |

**Table S3.** Fitting parameters for the kinetics shown in Figure S6(a) under back-side excitation at 560 nm. The decay dynamics were fitted using the multi-exponential function described by equation (S2).

| Phase | $\tau_1$ (ps) | $A_1$ (%) | $\tau_2$ (ps) | $A_2$ (%) | $\tau_3$ (ps) | $A_3$ (%) |
|---|---|---|---|---|---|---|
| n=2 | 0.12 | 70 (Decay) | 238 | 30 (Decay) | — | — |
| n=3 | 0.24 | 78 (Decay) | 1 | 4 (Rise) | 138 | 18 (Decay) |
| Bulk | 2.4 | 30 (Decay) | 16 | 30 (Rise) | 1000 | 40 (Decay) |

**Table S4.** Fitting parameters for the kinetics shown in Figure S6(b) under back-side excitation at 600 nm. The decay dynamics were fitted using the multi-exponential function described by equation (S2).

| Phase | $\tau_1$ (ps) | $A_1$ (%) | $\tau_2$ (ps) | $A_2$ (%) | $\tau_3$ (ps) | $A_3$ (%) |
|---|---|---|---|---|---|---|
| n=3 | 0.12 | 75 (Decay) | 111 | 25 (Decay) | — | — |
| Bulk | 1 | 52 (Rise) | 2000 | 48 (Decay) | — | — |

**Table S5.** Fitting parameters for the kinetics shown in Figure 5c under Front-side excitation at 750 nm. The decay dynamics were fitted using the multi-exponential function described by equation (S2).

| Phase | $\tau_1$ (ps) | $A_1$ (%) | $\tau_2$ (ps) | $A_2$ (%) | $\tau_3$ (ps) | $A_3$ (%) |
|---|---|---|---|---|---|---|
| n=1 | 17 | 26 (Rise) | 1000 | 74 (Decay) | — | — |
| n=2 | 0.3 | 97 (Rise) | 1000 | 3 (Rise) | — | — |
| n=3 | 0.3 | 4 (Rise) | 2 | 3 (Decay) | 2000 | 93 (Decay) |
| Bulk | 2 | 37 (Decay) | 1000 | 63 (Decay) | — | — |